\documentclass[prd,aps]{revtex4}

\usepackage{graphicx,subfigure}

\newcommand{\be}{\begin{equation}}
\newcommand{\ee}{\end{equation}}
\newcommand{\bea}{\begin{eqnarray}}
\newcommand{\eea}{\end{eqnarray}}
\newcommand{\bd}{\begin{displaymath}}
\newcommand{\ed}{\end{displaymath}}

\newcommand{\al}{\bar{A}}
\newcommand{\alb}{\bar{\alpha}}

\newcommand{\p}{ \partial_p }
\newcommand{\pp}{ \bar{\partial}_p }

\begin{document}

%\draft
%\preprint{
%\begin{tabular}{l}
%\hbox to\hsize{\hfill KAIST-TH 2006/10}\\
%[-1mm]
%\hbox to\hsize{\hfill KIAS-P06047}\\
%[-2mm] \hbox to\hsize{\hfill hep-ph/yymmdd}\\
%[-3mm] \hbox to\hsize{\hfill November 2011}\\
%[-3mm]
%\end{tabular}
%}

\title{
On the representation of the position operator and Lie-Hamilton equation in the discrete momentum space
}

\author{ Won Sang Chung }
\email{mimip4444@hanmail.net}

\affiliation{
Department of Physics and Research Institute of Natural Science, College of Natural Science, Gyeongsang National University, Jinju 660-701, Korea
}

\date{\today}

\begin{abstract}

In this paper the representation of the position operator and the Lie-Hamilton equation in the discrete momentum space.

\end{abstract}

\maketitle

\section{Introduction}

In 1992 Dimakis and Muller-Hoissen [1] used to non-commutative differential calculus to discuss the simple quantum mechanical system such as the one dimensional infinite square well potential. They dealt with one dimensional spatial lattice with finite spacing. That is to say, the space given in [1] is discrete.

About ten years ago,  E.Curado, M.Rego-Monteiro and H.Nazareno [2] discussed an analogous formalism for a momentum space instead of the position space. In many physical system, the momentum space is already discrete. For example, the allowed values for the momentum are positive integer in the one dimensional infinite square well potential. Thus, the non-commutative differential calculus is appropriate in dealing with the discrete momentum space.

Following the formalism in ref.[2], let us consider an one dimensional lattice in a momentum space where the momenta are allowed only to take discrete values , say $ p_0, p_0 + a , p_0 + 2 a , p_0 + 3 a , \cdots $, where $ a >0 $ is a spacing in a discrete momentum space. Let us denote the N-th momentum by $ p_N $  as follows :
\be
\{ p_N | ~ p_N = p_0 + ( N-1) a \}
\ee

The discrete momentum space appears in an one dimensional quantum mechanical system with zero potential between zero and $L$ and infinite elsewhere. It is well known that the energy is given by $E_N = \frac{ \hbar^2 \pi^2 }{2mL^2 } N^2 , ~N=1, 2, \cdots $. From the fact that $E_N = \frac{p_N^2}{2m} $, we see $ p_N = \frac { \hbar \pi }{L} N ,~N=1, 2, \cdots $. In this case we have $ p_0 = \frac { \hbar \pi }{L} , a = \frac { \hbar \pi }{L} $.

\section{Non-commutative differential calculus in the discrete momentum space}

The non-commutative differential calculus is based on the relation
\be
[ p, dp] = dp a ,
\ee
which induces the following
\be
f(p) dg(p) = dg(p) f(p+a)
\ee
for an arbitrary function $f$ and $g$.

Then we have two types of derivatives called the left derivative and the right derivative. The left derivative is defined by
\be
d f(p) = dp (  \p ) (p)
\ee
where
\be
( \p) (p) = \frac{1}{a} [ f( p+a ) - f(p) ]
\ee
The right derivative is defined by
\be
d f(p) =  (  \pp ) (p)dp
\ee
where
\be
( \pp) (p) = \frac{1}{a} [ f( p ) - f(p-a) ]
\ee
The Leibniz rule for the right derivative is given by
\be
(\p fg )(p) = (\p f ) (p) g(p) + f(p+a) (\p g ) (p)
\ee
and the Leibniz rule for the left derivative is given by
\be
(\pp fg )(p) = (\pp f ) (p) g(p) + f(p-a) (\pp g ) (p)
\ee
Introducing the momentum shift operators
\be
A = 1 + a \p , ~~~ \bar{A} = 1 - a \pp ,
\ee
which increase (decrease) the discrete momentum value by $a$
\be
( Af)(p_N) = f(p_{N+1} ) , ~~~ ( \bar{A} f ) (p_N) = f ( p_{N-1} )
\ee
and satisfies
\be
A \bar{A} = \bar{A} A =1
\ee

The momentum operator $P$ is defined as
\be
(Pf)(p) = p f(p)
\ee
for an arbitrary function $f(p)$ , where $P$ is hermitian. The the relation between $ A ( \bar{A} ) $ and $P$ is then given by
\be
[A, P ] = a A, ~~~ [ \bar{A} , P ] = - a \bar{A}
\ee
and the relation between $ A ( \bar{A} ) $ and the derivative is then given by
\be
[ \p , P ] = A, ~~~ [\pp , P ]= \al
\ee

The discrete integral called $ a $ -integral is defined as
\be
\int_{p_0}^{p_N} d_a p f(p) = a \sum_{j=0}^N f( p_0 + j a )
\ee
This integral is obtained by taking $ q \rightarrow 1 $ in the famous Hahn integral [3]. The inner product of two complex function $f$ and $g$ is defined as
\be
<f|g> =  \int_{p_0}^{p_N} d_a p f^* (p) g(p),
\ee
where $*$ means the complex conjugation. Then the function $f(p)$ with the inner product (17) spans the Hilbert space and the operator $ A$ and $ \bar{A} $ is defined as
\be
<f|Ag> = < \bar{A}f|g>,
\ee
which implies $ \bar{A} = A^{\dagger} $.
The eqs.(17) and (18) shows that $A$ is unitary operator. Thus, we can define a position operator $X$ as
\be
X = \frac {1}{2i} ( \p + \pp )
\ee

Now let us investigate the quantization procedure. To do so, we should compute the commutation relation between the position and momentum operator as follows :
\be
[X, P ] = -i + \frac {i} {2} a Q ,
\ee
where the operator $Q$ is given by
\be
Q = \pp - \p
\ee
If we define
\be
H = X^2 + P^2 ,
\ee
we have the following relation
\be
H = - \frac{1}{4a^2} ( A - \al )^2 + P^2
\ee
The Lie-Hamiltonian equation is then as follows :
\bd
[ X, H] = - 2 i P + i \frac{a}{2} \{ Q, P \} = -2i P + ia PQ + a^2 X
\ed
\be
[ P, H] =  2 i X - i \frac{a}{2} \{ Q, X \} = 2iX - ia XQ
\ee
This relation (24) reduces to the continuum case when $ a$ goes to zero.

\section{Representation of the position operator in the discrete momentum space}

In this section we will find the concrete form of the eigenvector $ \phi(p) $  for the position operator $X$ with eigenvalue $x$ as follows :
\be
X \phi(p) = x \phi (p),
\ee
where acting the position operator $P$ on $ \phi(p) $ gives
\be
P \phi(p) = p \phi(p)
\ee
Using the eq.(19), the eq.(25) is written by
\be
\phi( p+a) - \phi ( p-a) = 2iax \phi (p)
\ee
The functional relation (27) is written by
\be
\phi ( p+a) - \alpha \phi (p) = - \alb ( \phi(p) - \alpha \phi ( p-a) ) ,
\ee
where $\alpha $ is given by
\be
\alpha = iax - \sqrt{ 1-a^2 x^2 }
\ee
and $\alb$ is a complex conjugate of $ \alpha $ and we have
\be
\alpha \alb = 1
\ee

If we set
\be
h(p+a) = \phi ( p+a) - \alpha \phi(p),
\ee
we then have
\be
h(p+a) = - \alb h(p)
\ee
The recurrence relation (32) is solved as follows :
\be
h(p_N ) = ( -\alb )^N \phi(p_0)
\ee
where $ h(p_0 ) = \phi( p_0 ) $.
Because $ h(p_{N+1} ) = \phi ( p_{N+1} ) - \alpha \phi ( p_N ) $, we have
\be
\phi(p_N ) = \frac {\phi(p_0)}{\alpha + \alb } ( \alpha^{N+1} - ( -\alb)^{N+1} )
\ee

If we demand $ < \phi (p) | \phi (p) > =1 $ , we have
\be
|\phi(p_0 ) |^2 = \frac { ( \alpha + \alb )^2 }{a \left[ 2N +1 + \frac { (-\alpha^2 )^{N+1} - ( - \alpha^{-2} )^N } {\alpha^2 + 1 } \right] }
\ee

\section*{Conclusion}

In this paper we used non-commutative differential calculus to study one dimensional quantum mechanical system in the discrete momentum space. We investigated the quantization procedure for this model and studied the Lie-Hamiltonian equation. Moreover we found the concrete form of the eigenvector for the position operator in the momentum space.

%%%%%%%%%%%%%%%%%% References
%%%%%%%%%%%%%%%%%%%%%%%%%%%%%%%%%%%%%%%%%%%%%%%%%%%%%%%%%%%%%%%%%%%%%%%
\def\JMP #1 #2 #3 {J. Math. Phys {\bf#1},\ #2 (#3)}
\def\JP #1 #2 #3 {J. Phys. A {\bf#1},\ #2 (#3)}
\def\JPD #1 #2 #3 {J. Phys. D {\bf#1},\ #2 ( #3)}
\def\PRL #1 #2 #3 { Phys. Rev. Lett. {\bf#1},\ #2 ( #3)}

%%%%%%%%%%%%%%%%%%%%%%%%%%%%%%%%%%%%%%%%%%%%%%%%%%%%%%%%%%%%%%%%%%%%%%%

\section*{Refernces}
[1] A.Dimakis and F.Muller-Hoissen , Phys.Lett.B {\bf 295} 242 (1992).

[2] E.Curado, M.Rego-Monteiro and H.Nazareno, hep-th/00122444v2

[3] Hahn, W., Math.Nachr.{\bf 2 }, 4 (1949)

\end{document}